\documentclass[12pt]{iopart}

\usepackage{iopams}
\usepackage{amssymb}
\usepackage{graphicx}
\begin{document}

\title[Dark soliton-induced reshaping of quantum states of a dilute Bose gas]{Using dark solitons from a Bose-Einstein condensate necklace to imprint soliton states in the spectral memory of a free boson gas}

\author{Alain Mo\"ise Dikand\'e}

\address{laboratory of Research on Avanced Materials and Nonlinear Science (LaRAMaNS), Department of Physics, Faculty of Science, University of Buea PO Box 63 Buea, Cameroon}
\ead{dikande.alain@ubuea.cm}
\vspace{10pt}
\begin{indented}
\item[]July 2023
\end{indented}

\begin{abstract}
A possible use of matter-wave dark-soliton crystal produced by a Bose-Einstein condensate with ring geometry, to store soliton states in the quantum memory of a free boson gas, is explored. A self-defocusing nonlinearity combined with dispersion and the finite size of the Bose-Einstein condensate, favor the creation of dark-soliton crystals that imprint quantum states with Jacobi elliptic-type soliton wavefunctions in the spectrum of the free boson gas. The problem is formulated by considering the Gross-Pitaevskii equation with a positive scattering length, coupled to a linear Schr\"odinger equation for the free boson gas. With the help of the matter-wave dark soliton-crystal solution, the spectrum of bound states created in the free boson gas is shown to be determined by the Lam\'e eigenvalue problem. This spectrum consists of $\vert \nu, \mathcal{L} \rangle$ quantum states whose wave functions and energy eigenvalues can be unambiguously identified. Among these eigenstates some have their wave functions that are replicas of the generating dark soliton crystal. 
\end{abstract}

%
%
\submitto{\NJP}
%
%
%

\section{Introduction}
The idea that a high-intensity optical field propagating in a nonlinear optical device, could be used to trap and reshape a photon field of a relatively lower power, has been considered in several instances. This great deal of interest follows Manassah's proposal  \cite{man1,man2} that an ultrafast pulse with energy lower than that required to self-sustain "solitonic" profile in an optical medium, may preserve its
shape provided an intense co-propagating pump of a different
color (or wavelength) and a longer duration typical of soliton, is present. Subsequent to this seminal conjecture, as well as to experimental evidences reported by de la Fuente \etal \cite{dela1,dela2}, it has
become evident that solitons could play a major role in the process of light control by
light \cite{7,8}. Thus numerical simulations \cite{9,10} have demonstrated that due to their outstanding stability, and particularly their robustness upon collisions with other excitations both of the same types and of different natures (including continuous waves), solitons could represent highly reliable guiding structures for reconfigurable
low-power signals and re-routinable high-power pulses. Sequel to this reconfiguration is the possibility to construct soliton networks that can be used as nonlinear waveguide arrays, as
established in the landscape of recent experiments \cite{11,12,13,14,15,16} emphasizing the outstanding robustness of the periodic lattice
of optical solitons with a flexible (hence controllable) refractive index modulation depth and period that are induced all-optically. Such periodic optical-soliton lattice will generally
form in continuous nonlinear media, where they develop into
optically imprinted modulations with some effective refractive
index. Due to the tunable character of the induced nonlinear refractive
index as well as the flexible modulation period, a great
number of new opportunities for all-optical manipulations
of light can be envisaged since in this context, periodic optical soliton waveguides can operate in both weak and strong-coupling regimes depending on the depth of the refractive
index modulation.
Soliton transmissions based on
optically-induced waveguiding configurations with soliton crystals
are familiar in a wide range of modern communication technologies including
mode-Locked lasers \cite{17}, photonic media \cite{9,10,18} and photorefractive
semiconductors \cite{19}.\par
Analogous to the process of photon trapping and transfer with photons, a system of non-interacting cold atoms could be manipulated by an effective field created by photons or a Bose-Einstein condensate (BEC). Besides the possibility for such physical process that has been discussed in several contexts of electromagnetically-induced transparency in cold-atom systems \cite{fei1,dik1}, a fascinating scenario was recently proposed by Shaukat \etal \cite{shau1} who suggested to use dark solitons of a repulsive BEC to create quantum states (then identified to be "qubits") in a free boson gas interacting with the BEC. Interesting enough this proposal puts matter-wave excitations, and particularly matter-wave dark solitons produced in a BEC, on the same footing with light as a macroscopic quantum system with sound potential for application in quantum information processing. From a more general standpoint the dynamics of dark solitons in BECs has been widely investigated over the recent past \cite{43,44,44a}, and several aspects of their stability have been explored. The unequivocal evidence from these studies is the higher stability of dark solitons, be it within the framework of scatterings of two BECs or of a single BEC interacting with defects or impurities \cite{njp1,njp2,njp3,njp4,njp5}. Macroscopic quantum phenomena in BECs have equally been explored in the contexts of BECs confined in a potential. Of particulalr interest is the situation involving periodic potentials described by Jacobi elliptic functions \cite{decon1}, whose wave functions are well known to form a family of periodic soliton solutions. Note that the stability of these periodic nonlinear wavetrain structures was probed against collisions with small-amplitude excitations \cite{decon1}. It is believed that the periodic dark soliton structures obtained in ref. \cite{decon1} could be relevant in the design of periodic matter-wave gratings \cite{choi}, in addition to providing a suitable tool for the fabrication of logic gates consisting of periodic arrays of BECs imprinted within magnetic chips \cite{jose}.\par
In the present work we examine the possibility to use a matter-wave dark soliton crystal produced by a repulsive BEC, to imprint macroscopic quantum states in the spectral memory of a free boson gas. The study considers a mixture of two coupled one-dimensional ($1D$) boson gases of two different atomic species, one of which forms a BEC with repulsive interactions whereas the other is a non-interacting system of Bose atoms. In ref. \cite{shau1} a similar problem was addressed for a mixture of two infinitely long $1D$ boson systems, in which the matter-wave excitation of the repuslive BEC was a "$tanh$"-shaped dark soliton. In connection with this dark-soliton solution, the time-independent linear Schr\"odinger equation for the free boson system turned into an eigenvalue problem with a "$sech^2$" potential well. Localized modes of this eignevalue problem were postulated to represent quantum states associated either with a two-level (qubit) or tree-level (qutrit) systems spectrum, depending on the population of bound states composing the spectrum. In our study we shall focus on the bound-state spectrum for the eigenvalue problem induced by the periodic dark-soliton structure from the repulsive BEC with finite size, in the free boson system. However, unlike \cite{shau1}, here the discrete spectrum of the free boson system will emerge to be more rich and complex, given that the periodic dark soliton is a "crystal" of entangled "$tanh$" dark solitons \cite{rod1} and the correlations of centers of mass
of these entangled "$tanh$" modes broaden the internal-mode spectrum, promoting a multitude of degenerate states. \par In section \ref{sec2} we introduce the model and obtain the periodic-soliton solution for the repulsive BEC. In principle this specific nonlinear solution can be associated with two different contexts of dynamics of the BEC with repulsive interaction: either a BEC with ring geometry, or an infinitely long repulsive BEC for which the GP equation is solved with periodic boundary conditions. We believe that the first context is more physical and throughout the study we shall assume that our BEC is of finite length $L$. In section \ref{sec3} we reformulate the time-independent linear Schr\"odinger equation for the free boson gas, using the matter-wave dark soliton crystal obtained by solving the repulsive BEC. So doing we establish that for an arbitrary value of the coupling coefficient of the repulsive BEC to the free boson gas, the time-independent linear Schr\"odinger equation can be reduced to the Lam\'e eigenvalue problem of a general order \cite{t1}. Bound states of this eigenvalue problem are shown to form a spectrum of well defined $\vert \nu, \mathcal{L}\rangle$ quantum states involving $2\nu + 1$ soliton modes, whose eigenfunctions can be found analytically. The study ends with section \ref{sec4} which will be devoted to concluding remarks.
\section{The model and matter-wave periodic dark soliton solution for the repulsive BEC necklace}\label{sec2}
Consider a mixture of two systems of $1D$ boson particles of different atomic species $m_1$ and $m_2$. Particles of mass $m_1$ weakly mutually interact (i.e. weakly repel them and others) such that at very low temperature, the system is in a macroscopic quantum state characteristic of a BEC. The other component of the mixture is composed of boson particles of mass $m_2$ that do not interact them and others, but which interact with atoms of the BEC system. \par Denoting by $\psi_{\alpha}(x,t)$ the macroscopic wave function of the BEC system and by $\psi_{\beta}(x,t)$ the wave function of the free boson system, in mean-field approximation \cite{gp1} the dynamics of the mixture of BEC-free boson gas can be described by the coupled set of nonlinear partial differential equations:     
\begin{eqnarray}
 i\hbar \frac{\partial \psi_{\alpha}}{\partial t}&=&- \frac{\hbar^2}{2m_{\alpha}}\frac{\partial^2 \psi_{\alpha}}{\partial x^2} + g_{\alpha}\vert \psi_{\alpha} \vert^2\psi_{\alpha}, \label{eq1a} \\
 i\hbar \frac{\partial \psi_{\beta}}{\partial t}&=&- \frac{\hbar^2}{2m_{\beta}}\frac{\partial^2 \psi_{\beta}}{\partial x^2} + g_{\beta} \vert \psi_{\alpha} \vert^2 \psi_{\beta}. \label{eq1b}
\end{eqnarray}
In the above set $x$ and $t$ are space and time variables respectively. The coefficient $g_{\alpha}$ is the positive scattering length representing the strength of repulsion between particles in the BEC, and $g_{\beta}$ is a cross-coupling coefficient representing the strength of interaction of the repulsive BEC with the $1D$ free boson gas. In the following we shall assume $g_{\beta}$ to be positive. \par
Eq. (\ref{eq1a}) is the Gross-Pitaevskii equation \cite{gp1}, describing the spatio-temporal evolution of the BEC wave function $\psi_{\alpha}$. As well known \cite{gp1}, when the BEC system is infinitely long (i.e. $L\rightarrow \pm\infty$), the Gross-Pitaevskii equation (\ref{eq1a}) with positive scattering length admits a nonlinear solution so-called matter-wave dark soliton \cite{shau1}. This matter-wave dark soliton represents a deep in the BEC with a tail and front (of the deep) locking assymptotically onto two eventually degenerate constant amplitudes around the condensate edges $L\rightarrow \pm \infty$. Such a "smoothed-out" structure with a "kink" shape is mathemtically described by a "$tanh$" function, see for instance \cite{shau1}. \par
When the size $L$ of the BEC is finite, the "kink"-shaped soliton will not have the required physical environment to fully deploy. In this case the "$tanh$" soliton solution of the GP equation becomes unstable. However besides this hyperbolic kink-soliton solution, the Nonlinear Schr\"odinger equation with self-defocusing nonlinearity can also admit a periodic-soliton structure which is represented by a Jacobi elliptic function \cite{decon1,rod1,dik2,dik3}. This last periodic solution is the kind of soliton strucuture we are interested in, in connection with its periodic feature we can readily assume that it forms in a BEC with a ring shape of effective length $L$. Seeking for the periodic-soliton solution of the GP equation (\ref{eq1a}), we consider a wave function that describes a stationary state. Thus we express $\psi_{\alpha}(x,t)$ as the wave function of a quantum state with a standing amplitude $\phi_{\alpha}(x)$, undergoing temporal harmonic modulations at a frequency $\omega=E_{\alpha}/\hbar$, where $E_{\alpha}$ is the BEC energy:
\begin{equation}
 \psi_{\alpha}(x,t)=\phi_0(x)\,e^{-i\omega_{\alpha} t}. \label{s1}
\end{equation}
Substituting (\ref{s1}) in the GP equation (\ref{eq1a}) and performing a quadrature transformation, we obtain the following first-integral equation for the amplitude $\phi_{\alpha}(x)$ \cite{abram}: 
\begin{equation}
a\int{\frac{d\phi_{\alpha}}{\sqrt{(a^2-\phi_{\alpha}^2)(b^2 - \phi_{\alpha}^2)}}}=sn^{-1}\bigg(\frac{\phi_{\alpha}}{b}, \kappa\bigg)= ax\sqrt{\frac{\tilde{g}_{\alpha}}{2}}, \label{eqq5}
\end{equation}
where $sn()$ is the Jacobi Elliptic function of "snoidal" type with a modulus $\kappa$ \cite{abram}. Instructively, the modulus $\kappa$ of Jacobi Elliptic functions takes on values in the interval $ \leq \kappa \leq1$. The quantities $a$ and $b$ in the integral equation (\ref{eqq5}) are defined as:
\begin{eqnarray}
a^2&=&\frac{2}{1+\kappa^2}\frac{q^2}{\tilde{g}_{\alpha}}, \nonumber \\
b^2&=&a^2\kappa^2=\frac{2\kappa^2}{1+\kappa^2}\frac{q^2}{\tilde{g}_{\alpha}}, \nonumber \\ \tilde{g}_{\alpha}&=&\frac{2m_1g_{\alpha}}{\hbar^2}, \qquad q^2=\frac{2m_1 E_{\alpha}}{\hbar^2}. \label{ab}
\end{eqnarray}
Replacing these definitions in eq. (\ref{eqq5}) we obtain the following nonlinear solution:
\begin{eqnarray}
\phi_{\alpha}(x)&=&\phi_{0,\kappa}\,\,sn\bigg(\frac{x}{\ell_{\kappa}}\bigg), \label{ps1} \\
\ell_{\kappa}&=&\frac{\sqrt{1+\kappa^2}}{q}, \qquad \phi_{0,\kappa}=q\kappa\sqrt{\frac{2}{\tilde{g}_{\alpha}(1+\kappa^2)}}. \label{eqq7}
\end{eqnarray}
Remarkably, when $\kappa \rightarrow 1$ the snoidal function $sn()\rightarrow tanh()$ by virtue of asympotic properties of the Jacobi Elliptic functions. It is worth noting that the Jacobi Elliptic function $sn()$ is periodic with respect to its argument $x$. For the solution given by (\ref{ps1}) the period is $L=2\ell_{\kappa} K(\kappa)$, where $K(\kappa)$ is the complete Jacobi integral of first kind \cite{abram}. Interesting enough when $\kappa\rightarrow 1$, the quantity $K(\kappa)\rightarrow \infty$ and so $L\rightarrow \infty$ when the dark-soliton amplitude $\phi_{\alpha}(x)$ coincides with the hyperbolic-soliton solution to the Gross-Pitaevskii equation (\ref{eq1a}). 

\section{From linear Schr\"odinger to Lam\'e eigenvalue equation for the gas of free bosons}\label{sec3}
We turn to the linear Schr\"odinger equation (\ref{eq1b}) for the wave function of the free boson gas. We first set:
\begin{equation}
\psi_{\beta}(x,t)=\phi_{\beta}(x)e^{-i\omega_{\beta}\,t}, \qquad \omega_{\beta}=E_{\beta}/\hbar, \label{st1}
\end{equation}
consistently with the assumption of a stationary-state regime of motion for the BEC serving as pump system, where $E_{\beta}$ is the energy eigenvalue of the free boseon gas. Substituting the stationary-state ansatz (\ref{st1}) in eq. (\ref{eq1b}) and introducing a new space variable as $z=x/\ell_{\kappa}$, we obtain the eigenvalue problem:  
\begin{eqnarray}
0&=&\frac{d^2 \phi_{\beta}}{d z^2} + \bigg[h(\kappa) - \frac{4 m_2}{\hbar^2}\frac{g_{\beta}}{\tilde{g}_{\alpha}}\, \kappa^2sn^2(z)\bigg]\phi_{\beta}, \label{eig1} \\
h(\kappa)&=&\frac{m_2(1+\kappa^2)}{m_1}\frac{E_{\beta}}{E_{\alpha}}. \label{eiener}
\end{eqnarray}
Most generally the quantity $\frac{4 m_2}{\hbar^2}\frac{g_{\beta}}{\tilde{g}_{\alpha}}$ will be real and positive, in which case eq. (\ref{eig1}) is an eigenvalue equation for a "particle" of mass $m_2$ in a periodic lattice of "$sech^2$" potential wells. However, if for some selected values of characteristic parameters interfering this ratio, the quantity $\frac{4 m_2}{\hbar^2}\frac{g_{\beta}}{\tilde{g}_{\alpha}}$ is a positive integer, then we can set:
\begin{equation}
\frac{4 m_2}{\hbar^2}\frac{g_{\beta}}{\tilde{g}_{\alpha}}=\nu (\nu+1), \label{rat}
\end{equation}
Where $\nu$ is a positive integer determined by the last relation and taking values $\nu=0$, 1, 2, 3, ... With this the eigenvalue equation (\ref{eig1}) becomes:
\begin{equation}
\frac{d^2 \phi_{\beta}}{d z^2} + \bigg[h(\kappa) - \nu (\nu+1)\, \kappa^2sn^2(z)\bigg]\phi_{\beta}=0. \label{eig2}
\end{equation}
Eq. (\ref{eig2}) is nothing else but the Lam\'e equation of order $\nu$ \cite{t1}, for which we shall examine the bound-state spectrum for some selected values of the positive integer $\nu$. \par
The way the Lam\'e equation (\ref{eig2}) is derived, suggests that this equation can readily be regarded as the eigenvalue equation for quantum states created by the BEC in the spectral memory of the free boson gas. In fact the matter-wave soliton crystal (\ref{ps1}) creates an artificial potential that imprints quantum states in the spectrum of the free boson gas. For a given value of $\nu$ fixed by eq. (\ref{rat}), the discrete spectrum formed by these quantum states will consist of a set of $2\nu + 1$ modes. To explicitely determine how many of such modes are expected let us introduce a secondary quantum number $\mathcal{L}$, assuming that $\nu$ is the principal quantum number. It follows that for a given $\nu$, there will be $2\nu+1$ distinct modes $\vert \nu, \mathcal{L} \rangle$ denoted $\vert \nu, 1\rangle$, $\vert \nu, 2\rangle$, $\vert \nu, 3\rangle$, ..., $\vert \nu, \mathcal{L}=2\nu+1\rangle$. Let us examine some of these modes for two values of $\nu$ namely $\nu=1$ and $\nu=2$.
\subsection{Three-level system with basis states $\vert \nu=1, \mathcal{L}(=1, 2, 3)\rangle$:}
For $\nu=1$, the spectrum generated by the Lam\'e equation (\ref{eig2}) is composed of the three only allowed eigenstates of a three-level system which are:

\begin{itemize}
 \item Quantum state $\vert 1, 1\rangle$: 
 \begin{equation}
\phi_{\beta}^{(1)}(z)=\phi_{(1,1)} sn(z), \qquad E_{\beta,1}=\frac{m_1}{m_2}E_{\alpha}. \label{m11} 
 \end{equation}
 
\item Quantum state $\vert 1, 2\rangle$:
 \begin{equation}
\phi_{\beta}^{(2)}(z)= \phi_{(1,2)} cn(z), \qquad E_{\beta,2}=\frac{m_1}{m_2}\frac{E_{\alpha}}{1+\kappa^2}. \label{m12} 
 \end{equation}
 
 \item Quantum state $\vert 1, 3\rangle$:
 \begin{equation}
\phi_{\beta}^{(3)}(z)=\phi_{(1,3)} dn(z), \qquad E_{\beta,3}=\frac{m_1}{m_2}\frac{\kappa^2 E_{\alpha}}{1+\kappa^2}. \label{m13} 
 \end{equation}
 \end{itemize}
In the expressions of wave functions $\phi_{\beta}^{(i)}(z)$, the quantity $\phi_{(\nu,\mathcal{L})}$ is the normalization coefficient. \par
\begin{figure}\centering
\begin{minipage}[c]{0.5\textwidth}
\includegraphics[width=2.5in, height= 1.25in]{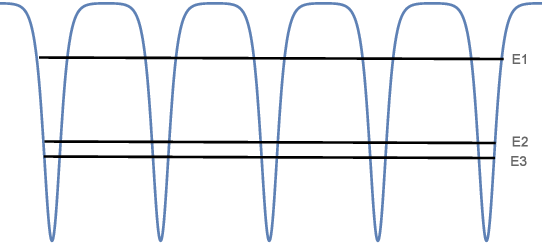}
\end{minipage}%
\begin{minipage}[c]{0.5\textwidth}
\includegraphics[width=2.5in, height= 1.12in]{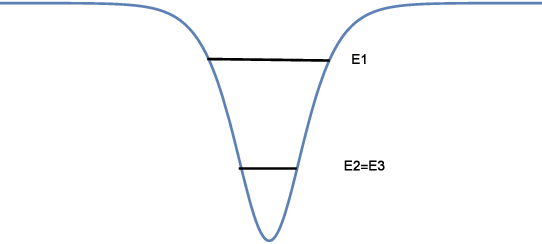}
\end{minipage}
      \caption{(Colour online) Distribution of the three energy levels $E_1\equiv \vert 1, 1\rangle$, $E_2\equiv \vert 1, 2\rangle$ and $E_3\equiv \vert 1, 3\rangle$, in the potential induced by the matter-wave soliton crystal from the repulsive BEC. Left graph: $\kappa=0.99$ (Periodic lattice of "$sech^2$" potential wells). Right graph: $\kappa=1$ (single "$sech^2$" potential well).}{\label{fig1}}
  \end{figure}

\begin{figure}\centering
\includegraphics{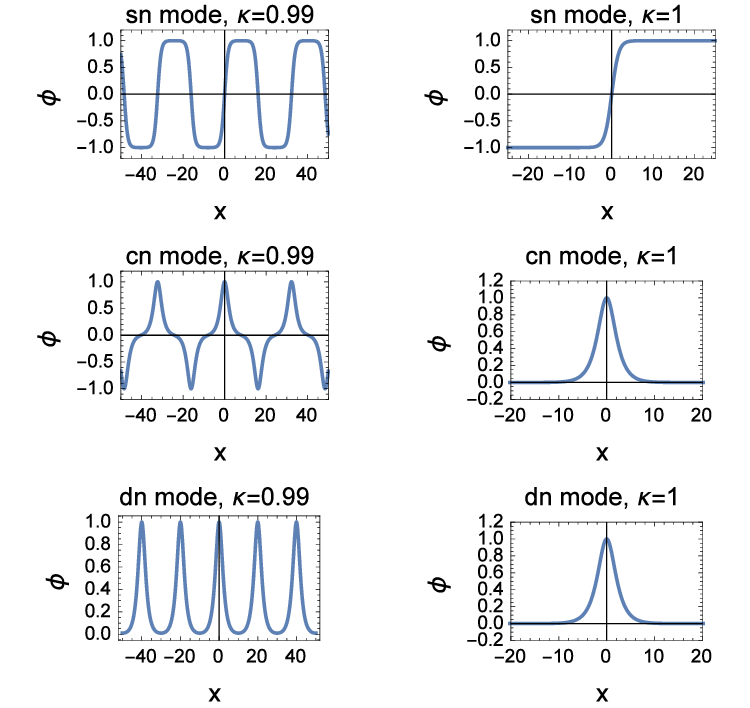}
\caption{(Colour online) Wave functions $\phi_{\beta}$ of the three bound states of the first-order Lam\'e equation, plotted versus $x$ for $\kappa=0.99$ (left column) and $\kappa=1$ (right column).}{\label{fig2}}
  \end{figure}

 What becomes the above three-level system when $\kappa\rightarrow 1$? In this limit $\phi_{\beta}^{(2)}(z)=\phi_{\beta}^{(3)}(z)\propto sech(z)$, and $E_{\beta,2}=E_{\beta,3}=(m_1/2m_2)E_{\alpha}$. This means that the two eigenstates $\vert 1,2\rangle$ and $\vert 1, 3\rangle$ merge into one single state at $\kappa= 1$, suggesting that near $\kappa=1$ these two states are closely degenerates. Because of this degeneracy the three-level system will look more like a two-level system, for which one of the two states is a compound of two closely degenerate states. The two-level nature of the system becomes well defined at $\kappa=1$, since in this limit the discrete spectrum will be composed of two well separated quantum state reminiscent of a qubit. Fig. \ref{fig1} is a sketch of the spectrum of eigenstates of the first-order Lam\'e equation, for $\kappa=0.99$ (left graph) and $\kappa=1$ (right graph). As we see, when $\kappa=0.99$ the potential created by the matter-wave dark soliton crystal is a periodic lattice of $sech^2$ potential wells. When $\kappa=1$, the periodic potential decays into a single $sech^2$ potential well as seen in the right graph of fig. \ref{fig1}.   
The state $\vert 1, 1\rangle$ has a wave function that mimics the profile of the mother soliton crystal (\ref{ps1}). We attribute this behavior to a clone of itself that the mother soliton crystal creates in the spectral memory of the free boson gas. This clone is the most energetically expensive to the dark soliton crystal of the three modes, as one can be convinced of by comparing the energy eigenvalues of the three states (\ref{m11}), (\ref{m12}) and (\ref{m13}). In table \ref{tab1} we summarize the expressions of wave functions and energy eigenvalues of the three modes when $\kappa=1$, while in fig.\ref{fig2} the three wave functions are plotted versus $z$ for $\kappa=0.99$ (graphs in the left column) and $\kappa=1$ (graphs in the right column).  
\begin{table}
\caption{\label{tab1} wave functions and energy eigenvalues of the five states composing the spectum induced by the soliton crystal for $\nu=2$, in the limit $\kappa=1$.}
\begin{tabular*}{\textwidth}{@{}l*{15}{@{\extracolsep{0pt plus
12pt}}l}}
\br
Eigenstate &Wave funtion&Energy\\
\mr
$\vert 1,1\rangle$&$tanh(x)$&$\frac{m_1}{m_2}E_{\alpha}$\\
$\vert 1,2\rangle$&$sech(z)$&$\frac{m_1}{2m_2}E_{\alpha}$\\
$\vert 1,3\rangle$&$sech(z)$&$\frac{m_1}{2m_2}E_{\alpha}$\\
\br
\end{tabular*}
\end{table} 

\subsection{Five-level system with basis states $\vert \nu=2, \mathcal{L}(=1, 2, 3, 4, 5)\rangle$:}
When $\nu=2$, the spectrum generated by the Lam\'e equation (\ref{eig2}) is composed of the five only allowed eigenstates of a five-level system which are:

\begin{itemize}
 \item Quantum state $\vert 2, 1\rangle$: 
 \begin{equation}
\phi_{\beta}^{(2,1)}(z)=\phi_{(2,1)} sn(z)cn(z), \qquad E_{\beta,2,1}=\frac{m_1(4+\kappa^2)}{m_2(1+\kappa^2)}E_{\alpha}. \label{m21} 
 \end{equation}
 
\item Quantum state $\vert 2, 2\rangle$:
 \begin{equation}
\phi_{\beta}^{(2,2)}(z)= \phi_{(2,2)} sn(z) dn(z), \qquad E_{\beta,2, 2}=\frac{m_1(1+4\kappa^2)}{m_2(1+\kappa^2)}E_{\alpha}. \label{m22} 
 \end{equation}
 
 \item Quantum state $\vert 2, 3\rangle$:
 \begin{equation}
\phi_{\beta}^{(2,3)}(z)=\phi_{(2,3)} cn(z)dn(z), \qquad E_{\beta,2,3}=\frac{m_1}{m_2}E_{\alpha}. \label{m23} 
 \end{equation}

\item Quantum state $\vert 2, 4\rangle$:
\begin{eqnarray}
\phi_{\beta}^{(2,4)}(z)&=&\phi_{(2,4)}\bigg[sn^2(z) -\frac{1+\kappa^2 + \sqrt{1-\kappa^2\kappa'^2}}{3\kappa^2}\bigg], \label{s24}  \\ E_{\beta,2,4}&=&\frac{m_1}{m_2(1+\kappa^2)}\bigg[2(1+\kappa^2)-\sqrt{1-\kappa^2\kappa'^2}\bigg]E_{\alpha}. \label{m24} 
 \end{eqnarray}
 
 \item Quantum state $\vert 2, 5\rangle$:
\begin{eqnarray}
\phi_{\beta}^{(2,5)}(z)&=&\phi_{(2,5)}\bigg[sn^2(z) -\frac{1 +\kappa^2 - \sqrt{1 -\kappa^2\kappa'^2}}{3\kappa^2}\bigg], \label{s25}  \\ E_{\beta,2,5}&=&\frac{m_1}{m_2(1+\kappa^2)}\bigg[2(1+\kappa^2)+\sqrt{1-\kappa^2\kappa'^2}\bigg]E_{\alpha}. \label{m25} 
 \end{eqnarray}
 \end{itemize}
 
\begin{figure}\centering
\begin{minipage}[c]{0.52\textwidth}
\includegraphics[width=2.5in, height= 1.25in]{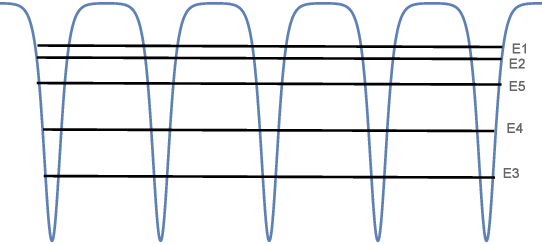}
\end{minipage}%
\begin{minipage}[c]{0.52\textwidth}
\includegraphics[width=2.5in, height= 1.25in]{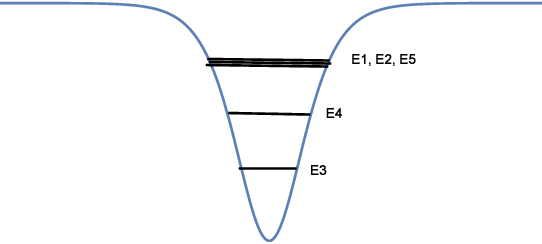}
\end{minipage}
      \caption{(Colour online) Five-level spectrum for the trapped boson gas when $\nu=2$: $E_1\equiv \vert 2, 1\rangle$, $E_2\equiv \vert 2, 2\rangle$, $E_3\equiv \vert 2, 3\rangle$,  $E_4\equiv \vert 2, 4\rangle$ and $E_5\equiv \vert 2, 5\rangle$, in the potential induced by the matter-wave soliton crystal from the repulsive BEC. Left graph: $\kappa=0.99$ (Periodic lattice of "$sech^2$" potential wells). Right graph: $\kappa=1$ (single "$sech^2$" potential well).}{\label{fig3}}
  \end{figure}

In fig. \ref{fig3}, we sketch the spectrum of five states of the Lam\'e equation for $\nu=2$. 
  
When $\kappa\rightarrow 1$, the two states $\vert 2, 1\rangle$ and $\vert 2, 2\rangle$ merge to form one single state with wave function $\phi_{\beta}^{(2,1)}(z)=\phi_{\beta}^{(2,2)}(z)\propto tanh(z)sech(z)$ and energy eigenvalue $E_{\beta,2,1}=E_{\beta,2,2}=(5/2)(m_1/m_2)E_{\alpha}$. In the same limit the two eigenstates  $\vert 2, 4\rangle$ and $\vert 2, 5\rangle$ remain completely separated, with wave functions $\phi_{\beta}^{(2,4)}(z)\propto - sech^2(z)$ and $\phi_{\beta}^{(2,5)}(x)\propto 2/3 - sech^2(z)$ and energy eigenvalues $E_{\beta,2,4}=(3/2)(m_1/m_2)E_{\alpha}$, $E_{\beta,2,5}=(5/2)(m_1/m_2)E_{\alpha}$. The third eigenstate $\vert 2, 3\rangle$ has its wave function that becomes $\phi_{\beta}^{(2,3)}(z)\propto sech^2(z)$ with the same energy eigenvalue $E_{\beta,2,3}=(m_1/m_2)E_{\alpha}$, when $\kappa\rightarrow 1$. Table \ref{tab2} summarizes the analytical expressions of the wave functions and energy eigenvalues of the five modes when $\kappa=1$ 
\begin{table}
\caption{\label{tab2} Wave functions and energy eigenvalues of the five states composing the discrete spectum induced by the soliton crystal for $\nu=2$, in the limit $\kappa=1$.}
\begin{tabular*}{\textwidth}{@{}l*{15}{@{\extracolsep{0pt plus
12pt}}l}}
\br
Eigenstate &Wave funtion&Energy\\
\mr
$\vert 2,1\rangle$&$tanh(x)sech(z)$&$\frac{5m_1}{2m_2}E_{\alpha}$\\
$\vert 2,2\rangle$&$tanh(x)sech(z)$&$\frac{5m_1}{2m_2}E_{\alpha}$\\
$\vert 2,3\rangle$&$sech(z)^2$&$\frac{m_1}{m_2}E_{\alpha}$\\
$\vert 2,4\rangle$&$-sech(z)^2$&$\frac{3m_1}{2m_2}E_{\alpha}$\\
$\vert 2,5\rangle$&$2/3 -sech(z)^2$&$\frac{5m_1}{2m_2}E_{\alpha}$\\
\br
\end{tabular*}
\end{table} 
 
\begin{figure}\centering
\includegraphics[width=5.8in, height= 6.7in]{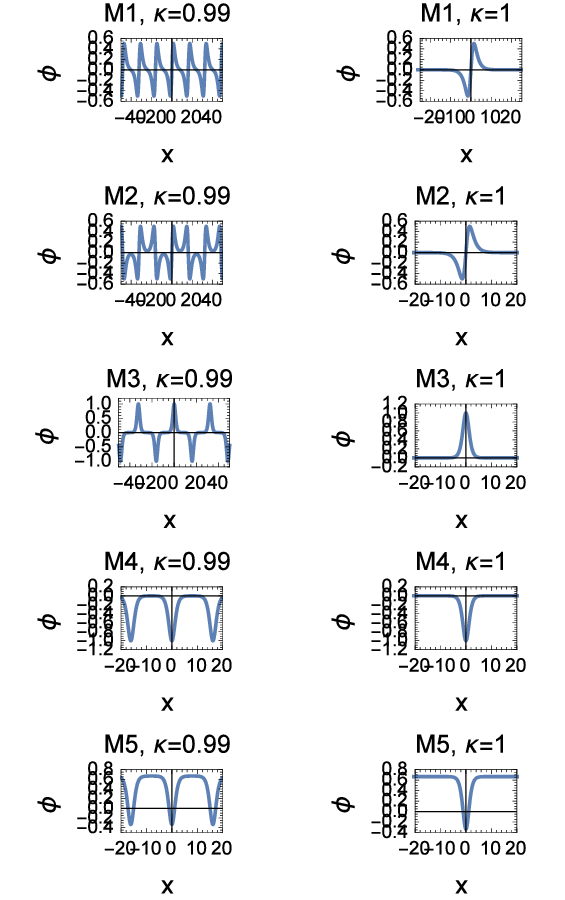}
\caption{(Colour online) Wave functions $\phi_{\beta}$ of the five bound states of the second-order Lam\'e equation, plotted versus $z$ for $\kappa=0.99$ (left column) and $\kappa=1$ (right column). The notations $M_1$, $M_2$, $M_3$, $M_4$ and $M_5$ in the graphs refer to the wave functions given respectively in (\ref{m21}), (\ref{m22}), (\ref{m23}), (\ref{s24}) and (\ref{s25}).}{\label{fig4}}
  \end{figure}

Table \ref{tab2} suggests that when $\kappa=1$, the discrete spectrum of eigenstates created in the free boson gas becomes a four-level spectrum. As we can see, two of these four eigenstates (namely the states $\vert 2,1\rangle =\vert 2,2\rangle$ and $\vert 2, 5\rangle$) have the same energy but with completely different wave functions. Also remarkable is the fact that the third state  $\vert 2, 3\rangle$ mimics the derive of the mother soliton crystal (\ref{ps1}), a behavior reminiscent of a "translational mode" of the mother soliton crystal with finite energy. The energy of this "energetic Golstone mode" is $\kappa$-invariant, which traduces its robustness. According to its expression, the energy of this "translational" mode is equal to the energy of the mother soliton crystal times the ratio of the two masses $m_1$ and $m_2$. In fig. \ref{fig4} we represent wave functions of the five modes corresponding to $\nu=2$, for $\kappa=0.99$ and $\kappa=1$. 

  \section{Summary and concluding remarks}\label{sec4}
  
Progress in quantum computing has led to the emergence of new concepts in the theory of information encoding, some among the most relevant being the concepts of quantum bits (qubits), quantum trits (quitrits) and so on. These concepts draw their importance from the fact that in quantum information theory, information is encoded in quantum states of particles such as electrons, atoms or photons. More precisely the concepts of quibit or qutrit describe the intrinsic nature of a specific quantum information, and can consist of distinct states corresponding to allowed quantum representations of some $n$-level system. These quantum states form a spectrum with well defined characteristic properties, namely a specific population as well as well defined characteristic features. Physical systems so far identified as being ideal candidates for quantum information processing include photonic crystals, superconducting nanowires and cold-atom systems \cite{ca1,ca2} including BECs. Concerning the latter physical system, it was recently suggested that a matter-wave dark soliton generated in a repulsive BEC, could be used to manipulate quantum states of a free boson gas \cite{shau1}. However, prior to this suggestion, it has been well established both theoretically \cite{man1,man2} and experimentally \cite{dela1,dela2} that light could be an excellent medium to trap and convey a light field of different mode (i.e. colour and power). In refs. \cite{9,10,18,dik2,dik3} for instance it was shown that bright and dark solitons, conveyed by an optical fiber or a photonic crystal, could be used to trap and reconfigure a photon field of lower power that may probably not survive when propagating lonely. In view of the current great interest to the issue of quantum information processing with cold-atom systems \cite{fei1,dik1}, it seemed to us an evident subject of interest to explore the possibility to use ideas suggested for photon trapping in quantum systems \cite{dik2,dik3}, for an understanding of the process in cold-atom systems. In this respect, in the present study we considered the specific physical context of a free boson gas driven by a BEC. In this context the BEC consists of mutually repelling atoms of identical mass $m_1$, resulting in a low-temperature dynamics described in mean-field approximation by a Gross-Pitaevskii equation with a positive scattering length. Assuming the BEC to be of finite length $L$, the application of periodic boundary conditions favored soliton solutions represented by the Jacobi Elliptic function of the snoidal type. In ref. \cite{rod1}, we established that the snoidal function is a "crystal" of "$tanh$-shaped kink-antikink" solitons. Using the matter-wave dark soliton crystal found, the linear Schr\"odinger equation for the free boson gas was shown to reduce to the Lam\'e equation of an arbitrary order $\nu$. Values of $\nu$, which is an integer determining the number of modes created by the BEC in the quantum memory of the free boson gas, appeared to be determined by the strength of coupling of the BEC to the free boson gas, compared with the nonlinearity coefficient of the BEC system.\par
In this study we considered the context of a BEC with repulsive interaction, coupled to a gas of free boson atoms. One could as well think of a context involving a repulsive BEC coupled to free fermion gas, or the case of a fermionic BEC coupled to a free boson gas, and so on. Also, in ref. \cite{dik3} we investigated quantum states of a photon trapped in the field created by a dark-bright optical soliton. In this work we found that the nature, namely shape profiles of the wave functions and values of the energy eigenvalues for the trapped photon, were determined by the competition between the two sub-systems generic of the dark and the bright soliton crystal structures. Such analysis would also be of valuable interest in the context of quantum computing involving cold-atom systems.  

\section*{Data availability statement}
No data were generated or analyzed in the present research.

\end{document}